\documentclass{article}

\usepackage{amsmath}
\usepackage{amssymb}
\usepackage{amstext}
\usepackage{bigstrut}
\usepackage{bm}
\usepackage{booktabs}
\usepackage[bf,footnotesize]{caption}
\usepackage{chngpage}
\usepackage{color}
\usepackage{dcolumn}
\usepackage{float}
\usepackage{fullpage}
\usepackage{graphicx}
\usepackage{lineno}
\usepackage{makecell}
\usepackage{mathpazo}
\usepackage{mathrsfs}
\usepackage{mathtools}
\usepackage{multirow}
\usepackage{rotating}
\usepackage{setspace}
\usepackage{subfigure}
\usepackage{threeparttable}
\usepackage{xr}
\usepackage{algorithm}
\usepackage{algorithmic}
\usepackage{upgreek}

\definecolor{darkred}{rgb}{0.75,0,0}
\definecolor{darkgreen}{rgb}{0,0.5,0}
\definecolor{darkblue}{rgb}{0,0,0.75}
\definecolor{darkorange}{rgb}{1,0.9,0.1}
\definecolor{dark}{rgb}{0,0,0}

\usepackage{caption}
\newcommand*\patchAmsMathEnvironmentForLineno[1]{%
	\expandafter\let\csname old#1\expandafter\endcsname\csname #1\endcsname
	\expandafter\let\csname oldend#1\expandafter\endcsname\csname end#1\endcsname
	\renewenvironment{#1}%
	{\linenomath\csname old#1\endcsname}%
	{\csname oldend#1\endcsname\endlinenomath}}%
\newcommand*\patchBothAmsMathEnvironmentsForLineno[1]{%
	\patchAmsMathEnvironmentForLineno{#1}%
	\patchAmsMathEnvironmentForLineno{#1*}}%
\AtBeginDocument{%
	\patchBothAmsMathEnvironmentsForLineno{equation}%
	\patchBothAmsMathEnvironmentsForLineno{align}%
	\patchBothAmsMathEnvironmentsForLineno{flalign}%
	\patchBothAmsMathEnvironmentsForLineno{alignat}%
	\patchBothAmsMathEnvironmentsForLineno{gather}%
	\patchBothAmsMathEnvironmentsForLineno{multline}%
}

\makeatletter

\allowdisplaybreaks

\begin{document}

\title{Dynamical heterogeneity reverses structural suppression of cooperation
}

\author{Xiaochen Wang$^{1,\ast}$
\\
\footnotesize{$^{1}$Center for Systems and Control, Peking University, Beijing 100871, China}\\
\footnotesize{$\ast$ Corresponding author. E-mail: xiaochen\_@pku.edu.cn}\\
}
\date{}
\maketitle

\begin{abstract}
Heterogeneity in individual characteristics and behaviour is a fundamental property of complex dynamical systems. 
While previous studies on evolutionary dynamics of strategies evolution in various systems have predominantly focused on the structural heterogeneity, dynamical heterogeneity in individuals' strategy update has been largely neglected. 
Here, we introduce a novel dynamical update mechanism based on individuals' decision-making information, comprising personal and social components. 
This update rule allows each individual to vary in the weight of personal information and the amount of social information, capturing the general scenario of dynamically heterogeneous populations.
We find that cooperation, as a collective prosocial outcome, is significantly enhanced when highly connected individuals on interaction network rely more heavily on personal information and access more social information.
This effect is notably absent in homogeneous networks, thereby overturning the prevailing consensus that structural heterogeneity inherently suppresses cooperation.
This theoretical prediction is further validated by empirical evidence from GitHub collaboration networks.
Furthermore, individuals preferentially linking to those who are well-informed and possess greater personal information further promotes collective cooperation. 
We additionally reveal that cooperators gain a decisive advantage when relying more on personal information compared to defectors, whereas social information affects cooperators and defectors equivalently. 
Our findings offer profound insights into how dynamical heterogeneity fundamentally shapes the evolution of collective cooperation in complex systems.
\end{abstract}

\section*{Introduction}
Heterogeneity is a ubiquitous feature of many real-world systems, shaping the structure and interactions across diverse domains, from biological ecosystems to social and technological networks \cite{mcdonnell2011benefits,watts1998collective,barabasi2009scale}. 
In many complex systems, individuals or components differ in their connectivity, behaviour, or local environment, leading to profound implications for collective dynamics and evolutionary outcomes. 
A prominent form of heterogeneity arises in the structure of interaction networks themselves \cite{barabasi1999emergence,watts1998collective}. 
Unlike regular lattices or fully connected graphs, real-world networks—such as social networks, neural networks, and ecological food webs—often exhibit heterogeneous degree distributions \cite{wasserman1994social,sporns2004organization,dunne2002food,barabasi2009scale}, with some nodes being highly connected while others have only a few links. 
This structural heterogeneity has been shown to significantly influence dynamical processes such as disease spreading, opinion formation, and evolutionary dynamics \cite{albert2002statistical,newman2003structure,boccaletti2006complex}.
Specifically, heterogeneous network structures can significantly influence the evolutionary dynamics of group game interactions, particularly the emergence of group-optimal strategies such as cooperation.

In particular, heterogeneous networks have profound implications for evolutionary dynamics \cite{santos2005scale,allen2017evolutionary,mcavoy2020social}, which explains how cooperation emerges in systems. 
Compared to regular networks, heterogeneous networks can either suppress cooperation \cite{allen2017evolutionary} or enable the emergence of influential hubs and stable cooperative clusters \cite{santos2005scale}. 
However, despite numerous studies exploring the effects of heterogeneity on evolutionary dynamics, the existing literature predominantly focuses on network heterogeneity, largely neglecting another critical dimension: dynamical heterogeneity. 
This heterogeneity refers to the fact that each individual follows a distinct dynamical rule in strategy update. 
Specifically, in the context of evolutionary dynamics, this means that each individual follows their own unique update rule, which guides how they revise their strategy.

\begin{figure}[h]
	\centering
	\includegraphics[width=1\textwidth]{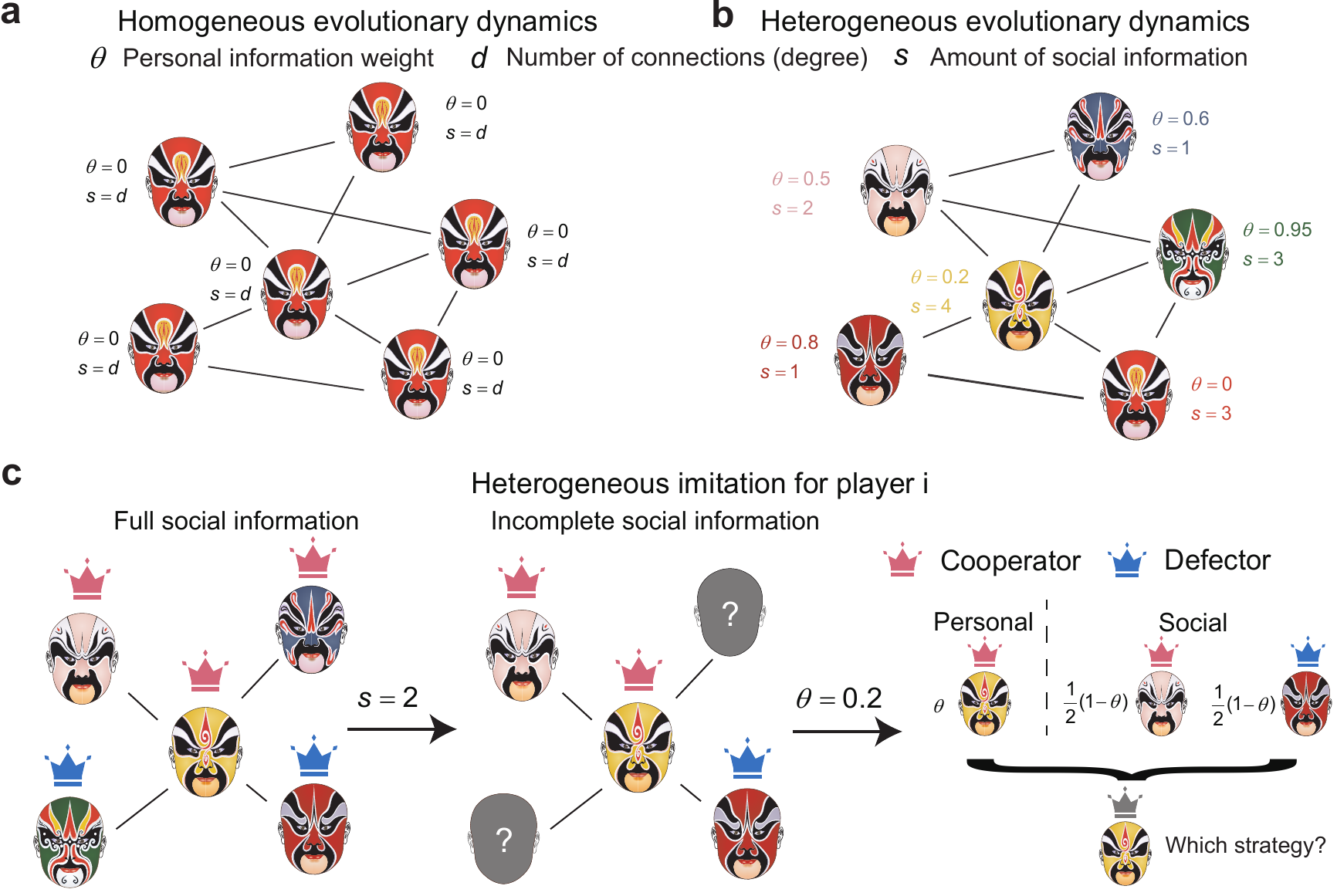}
	
	\caption{
    \textbf{Illustration on heterogeneous evolutionary dynamics.}
		\textbf{a},
		Under homogeneous evolutionary dynamics, each individual follows the same decision-making mechanism (illustrated as same coloured Peking opera masks).
        In the canonical death-birth updating, individual omit their personal information ($\theta=0$) and take all the neighbours' information into consideration ($s=d$).
		\textbf{b},
		Under heterogeneous evolutionary dynamics, each individual may follow a distinct decision-making mechanism (illustrated as different coloured Peking opera masks). Owing to their unique characteristics, each individual possesses a specific personal information weight $\theta$ and a particular amount of social information $s$.
		\textbf{c},
        Under heterogeneous imitation with incomplete social information, each individual obtains information about their neighbours’ strategies and payoffs based on their own amount of social information $s$, and then makes a decision and thus imitates one of them by weighting this social information against their own experience, according to their personal information weight $\theta$.
        Each individual on the network can have its own distinct personal information weight and amount of social information.
	}
	\label{fig: 1}
\end{figure}

Indeed, Most existing studies have focused on a single updating rule, such as death-birth or pairwise comparison \cite{ohtsuki2006simple,ohtsuki2007evolutionary,wang2023imitation,allen2017evolutionary}, assuming that differences in individuals' decision-making processes arise solely from variations in their number of connections (Fig.~\ref{fig: 1}a). 
A smaller number of studies have explored systems involving two or three updating rules \cite{cardillo2010co,takesue2019effects,you2020effects}, including scenarios where updating rules coevolve with strategies, or different update retes \cite{meng2024dynamics}. 
However, empirical systems are unlikely to be constrained to a limited set of updating mechanisms or variation in update rates. 
For example, in a celebrities social network \cite{barabasi1999emergence}, different individuals have access to varying amounts of information, and their openness to changing strategies also differs, which leads to personalized update mechanisms. 
Previous studies have largely overlooked this highly prevalent phenomenon.
In reality, each individual may possess a unique updating process, highlighting the necessity of considering broader dynamical heterogeneity.

Building upon this understanding of individual heterogeneity, recent study on imitation dynamics with incomplete information have opened new avenues for exploring dynamical heterogeneity \cite{wang2023imitation}. 
In this model, individuals base their decisions on a combination of personal information and external social information (i.e., information from their neighbors). 
Individuals can choose to utilize varying portions of social information and adjust the weighting between personal and social information in their decision-making processes. 
Past studies have mainly focused on homogeneous networks where each individual follows identical dynamical rules, which limits their applicability to a narrow set of scenarios. 
Here, we focus on scenarios with heterogeneous dynamics(Fig.~\ref{fig: 1}b), similar to those observed in real-world systems.
By allowing individuals to differ in how they process and prioritize information, these models provide a more realistic representation of decision-making processes in complex systems.

Here, we introduce an update rule called heterogeneous imitation (HIM), which allows individuals to differ both in the weighting of their personal information and in the amount of social information they utilise (Fig.~\ref{fig: 1}c). 
We demonstrate that cooperation is significantly promoted when high-degree nodes (hubs) access more social information and place greater emphasis on personal information—a discovery we empirically validate using GitHub collaboration networks. 
Moreover, our results reveal that during network formation, individuals preferentially connecting to those who are well-informed and possess greater personal information weights further promotes collective cooperation. 
Furthermore, when strategies and information coevolve, cooperation is greatly enhanced if cooperators rely more on personal information than defectors, whereas social information affects both types equivalently. 
Our study offers profound insights into how dynamical heterogeneity fundamentally shapes the evolution of collective cooperation in complex systems.

\clearpage

\section*{Model}

\subsection*{Game and payoff}
We consider a population of $N$ individuals, where interaction relationships are represented by an undirected and unweighted network.
At each time step, individual $i$ interacts with its $d_i$ neighbours on the network and receives payoffs from the game.
We focus on the donation game, in which each individual chooses to either cooperate (C) or defect (D).
Cooperators pay a cost $c$ to provide a benefit $b$ to each of their neighbours, while defectors do nothing.
In this model, mutual defection is the Nash equilibrium, while mutual cooperation is the more desirable outcome, which is called a social dilemma.

After interaction, each individual $i$’s payoff, $u_i$, is calculated as the average over its $d_i$ interactions.
This payoff is then transformed into the individual’s fitness by $F_i = \exp{(\delta u_i)}$, where $\delta > 0$ denotes the intensity of selection.
We consider the weak selection limit by assuming $\delta \to 0$, meaning that the contribution of game interactions to fitness is minimal.

\subsection*{Heterogeneous imitation dynamics}

After interactions, individuals update their strategies according to certain update rule.
Building on previous work \cite{wang2023imitation}, we propose a novel framework called ``heterogeneous imitation” (HIM).
Unlike all previous imitation-based update rules \cite{ohtsuki2006simple,allen2017evolutionary,wang2023imitation}, this update rule allows each individual to have their own independent update mechanisms.
Under HIM, the probability that an individual $i$ imitates one of its neighbours individual $j$ is
	\begin{equation}
		p_{i\to j} = \left\{
		\begin{aligned}
			&\frac{(1-\theta_i)/s_iF_j}{(1-\theta_i)/s_i\sum_{l\in \Omega^{(s_i)}_i }F_l+\theta_i F_i},\quad j\in \Omega^{(s_i)}_i,\\
			&\frac{\theta_i F_j}{(1-\theta_i)/s_i\sum_{l\in \Omega^{(s_i)}_i } F_l+\theta_i F_i},\quad  j=i,
		\end{aligned}
		\right.
		\label{eq:pij}
	\end{equation}
where $s_i\in\left\{1,\dots ,d_i\right\}$ is the number of neighbours' information consulted by individual $i$, and $\Omega^{(s_i)}_i$ denotes the set of $s_i$ neighbours that it randomly selects.
Another parameter, $\theta_i \in [0,1)$, known as the weight of personal information, represents the relative importance of personal information compared to social information in the process of updating the strategy.
For instance, individual $i$ may completely ignore its personal information ($\theta_i=0$), or it may rely on it heavily ($\theta_i\to1$).
If $\theta_i=0$ and $s_i=1$ for all individual $i$, the process is equivalent with the so-called random drift, where individuals imitate one of its neighbours strategy randomly and game takes no effect.
The vectors of social information $\mathbf{s}=(s_1,s_2,...,s_N)$ and personal information $\bm{\uptheta}=(\theta_1,\theta_2,...,\theta_N)$ together determine the information status of the entire population.

Such microscopic dynamics of evolution induces a Markov chain, which eventually ends in two absorbing states, namely, all-C or all-D, where all individuals share the same strategy.
We consider the fixation probability for both cooperators, $\rho_C$, and defectors, $\rho_D$, which represents the probability that a single mutant of a strategy invades and replaces the population of another strategy.
If $\rho_C>\rho_D$, cooperators are favoured relative to defectors.
For the donation game under weak selection, this condition corresponds to a critical value of the benefit-to-cost ratio, denoted as $(b/c)^*$.
If the condition corresponds to $b/c > (b/c)^* > 0$, cooperators will be favoured under such condition.
However, if it corresponds to $b/c < (b/c)^* < 0$, it implies that such a strategy can evolve only when a “cooperator” pays a cost $c$ to inflict a negative harm $b$ on their opponent.
We refer to this type of strategy as spite, which is a even worse outcome than defection.
Our goal is to promote collective cooperation and inhibit spite behaviour.

\section*{Results}

\begin{figure}[h]
		\centering
		\includegraphics[width=1\textwidth]{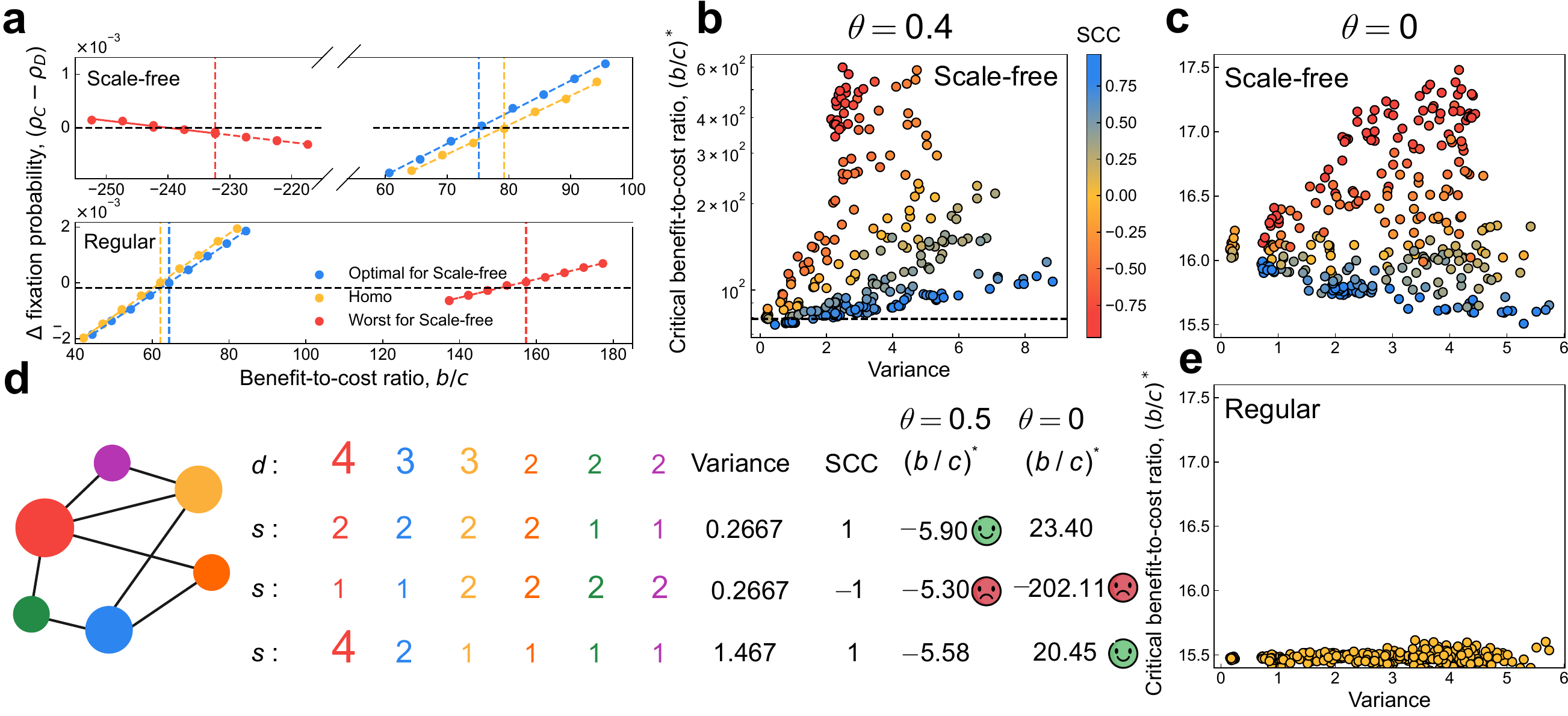}
		
		\caption{
			\textbf{Cooperation evolution under heterogeneous social information.}
			\textbf{a},
			We plot the fixation probability difference, $\rho_C-\rho_D$ as a function of benefit-to-cost ratio, $b/c$ under different distributions of social information.
			We generate three distributions of the amount of social information with the same average amount of social information $\langle s\rangle$ using the algorithm we developed.
			We find that cooperation can be promoted by the Optimal information situation, while inhibited by the Worst situation.
			We apply exactly the same distribution on the regular network, and find that homogeneous information best promotes cooperation, which is contrary to the situation in heterogeneous networks.
			We generate seiries of social information with different variance and Spearman correlation coefficient with $\mathbf{d}$ (SCC) on scale-free network (\textbf{b, c}) and regular network (\textbf{e}).
			On the scalre-free network, when $\theta$ is large, increasing variance and decreasing SCC can inhibit cooperation (b).
			And when $\theta=0$, increasing the variance can promote cooperation (c).
			However, on regular network, the difference of the critical benefit-to-cost ratio with distinct information situation is insignificant (e).
			\textbf{d},
			We generate a small heterogeneous network as an intuitive example.
			We give three information series: the first has small variance and prositive SCC, the second has small variance and negative SCC, and the third has large variance and positive SCC.
			We find that when $\theta$ is large, the first series best promote cooperation while the second one inhibit cooperation most.
			However, when $\theta=0$, the third series with largest variance can promote cooperation compared with small variance counterpart.
		}
		\label{fig: 2}
	\end{figure}
\subsection*{Heterogeneous social information}
We will explore how heterogeneous social information and personal information affect the emergence of cooperation separately.
First, we examine the influence of heterogeneous social information.
We assume that individuals' personal information remains the same (i.e., $\theta_i=\theta$ for all individual $i$), and study the impact on cooperation by varying the amount of social information available to them.

The amount of social information is closely tied to an individual's degree, as the latter serves as an upper bound for the former.
The greater the overall amount of social information in the population, the more favourable it is for cooperation \cite{wang2023imitation}. 
Therefore, we are interested in understanding which patterns of information allocation across nodes with different number of connnections (node degree) are most conducive to cooperation, given a fixed total amount of social information in the network.

To further shed light on the impact of information, we provide the criterion under which cooperation can outperform defection, which reads:
\begin{equation}\label{eq:competition}
\Bigg\langle \sum_{i\in G} \pi_ix_i
		\left\{
		\sum_{j\in G}
		p_{ij}^{(1)}\xi_{ij}^{(1)}
		\left(u_i-u_j\right)
		+\sum_{j\in G}
		p_{ij}^{(1)}\xi_{ij}^{(2)}
		\left(u_i-\sum_{l\in G}p_{jl}^{(1)}u_l\right)
		\right\} \Bigg \rangle^\circ>0,
\end{equation}
where $\langle\cdot\rangle$ denotes the average taken over random drift.
The parameter $p_{ij}^{(k)}$ means the probability that a $k$-step random drift starting from individual $i$'s position ends in individual $j$'s position.
The competition rate $\xi_{ij}^{(1)} = (1 - \theta_i)\left(\theta_i + \theta_j\right)$ captures the intensity of competition between first-order neighbours induced between individual $i$ and $j$,  
while $\xi_{ij}^{(2)} = (1 - \theta_i)(1 - \theta_j)\frac{d_j(s_j - 1)}{s_j(d_j - 1)}$ represents the competition intensity between second-order neighbours.

Equation.~(\ref{eq:competition}) implies that only by winning the competition against both first-order and second-order neighbours can a cooperator survive and evolve. 
However, the first-order neighbours of a cooperator will always have an advantage over the cooperator, as they benefit from the cooperator's donation. 
This makes winning first-order competition extremely difficult. 
In contrast, success in second-order competition can convert the cooperator's first-order neighbours into cooperators as well, promoting the aggregation of cooperators and thereby increasing their payoffs. 
Thus, second-order competition is key to the evolution of cooperation \cite{wang2023imitation}.

Therefore, to promote cooperation, we should maximise the proportion of second-order competition within the overall competitive interactions.
Next, we focus on finding the best and worst series of social information for evolution of cooperation.
We now have the optimization problem 
\begin{equation}
		\begin{aligned}
			\max_{\mathbf{s}}&\sum_{i\in G}\sum_{j\in G}w_{ij}\frac{\xi_{ij}^{(2)}}{\xi_{ij}^{(1)}+\xi_{ij}^{(2)}},
			\\
			s.t. &\quad \sum_{i\in G}s_i=N\overline{s},\\
			&\quad 1\leq s_i \leq d_i,\\
			&\quad \mathbf{s}\in\mathbb{Z}^n,
		\end{aligned}
		\label{eq:bests}
	\end{equation}
where $\overline{s}$ is the average of the amount of social information.
The adjacency weight $w_{ij}=1$ if individual $i$ and $j$ are linked, otherwise, $w_{ij}=0$.
By solving the problem defined by Eq.~(\ref{eq:bests}) and a counter problem (see SI for details), we obtain the optimal and worst social information configurations. 
It can be observed in Fig.~\ref{fig: 2}a that, on heterogeneous networks, the optimal configuration reduces $(b/c)^*$, while the worst will inhibit cooperation and even promote spite. 
On the contrary, if we transfer these social information distributions to a regular network, a homogeneous distribution proves to be the best for cooperation. 
This echoes the conclusions of previous studies \cite{wang2023imitation} and highlights the stark contrast between heterogeneous and homogeneous networks.

We now investigate which types of sequences are more conducive to promoting cooperation.
By examining the optimal sequences generated by solving the above problem, we identify two key characteristics: first, the amount of social information for each individual is proportional to their degree; second, the variance of the sequence is relatively low. 
Based on this, we generate a range of social information sequences and evaluate each using two metrics: variance and Spearman correlation coefficient (SCC) with the degree sequence. 
The SCC measures the correlation between two sequences: the closer it is to 1, the more positively correlated they are; the closer it is to -1, the more negatively correlated they are.
After calculating the critical benefit-to-cost ratio under different social information sequences, we find that when the weight of personal information $\theta$ is relatively large, sequences with low variance and positive correlation with degree most effectively promote cooperation (Fig.~\ref{fig: 2}b). As variance increases or the correlation between the social information sequence and the degree sequence becomes more negative, cooperation becomes more difficult to evolve. 
However, when the personal information weight $\theta$ is small, sequences with higher variance and a positive correlation with degree better facilitate cooperation (Fig.~\ref{fig: 2}c). 
This suggests that, on heterogeneous networks, nodes with higher degrees should be assigned greater amounts of personal information.
In contrast to heterogeneous networks, regular networks show no significant difference in cooperation levels across sequences with different variances (Fig.~\ref{fig: 2}e), which is consistent with previous findings \cite{wang2023imitation}.

To provide a clear perspective on this phenomenon, we present an illustrative example. 
In Fig.~\ref{fig: 2}d, we show a heterogeneous network with six individuals, arranged in descending order of degree.
We assign three different social information sequences to the nodes.
The first sequence is positively correlated with degree and has low variance; it most effectively promotes cooperation when personal information weight is large ($\theta = 0.5$). 
The second sequence is negatively correlated with degree and consistently suppresses cooperation the most among the three, across all settings. 
The third sequence is also positively correlated with degree but has high variance; it proves most effective in promoting cooperation when $\theta = 0$.

\subsection*{Heterogeneous personal information}

	\begin{figure}[h]
		\centering
		\includegraphics[width=1\textwidth]{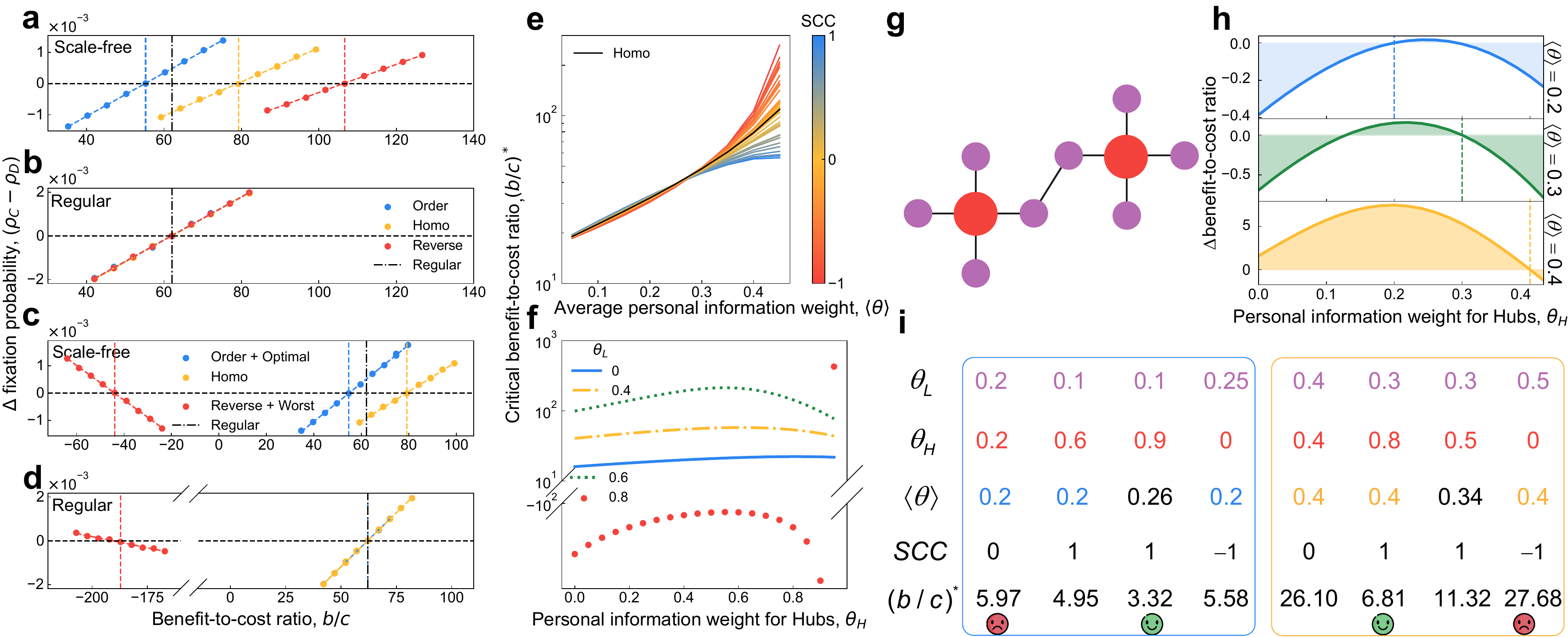}
		
		\caption{
			\textbf{Cooperation evolution under heterogeneous personal information.}
			We plot the fixation probability difference, $\rho_C-\rho_D$ as a function of benefit-to-cost ratio, $b/c$ under different distributions of personal information.
			We generate tree distributions of the amount of social information with the same average amount of social information $\langle s\rangle$ using the algorithm we developed.
			In the Order case, a larger degree $d_i$ for a individual corresponds to larger personal information weight $\theta_i$.
			And in the Reverse case, individuals with smaller degree get larger $\theta_i$.
			And in the Homo case, the personal information weights are identical.
			We find that the Order case strongly promote cooperation and the Reverse case inhibit cooperation on the scale-free network (a).
			However, there is no difference between the three cases on the regular network (b).
			The combination of different personal and social information distribution can examplify the phenomena.
			\textbf{e},
			We generate a series of personal information uniformly, and then randomize it to get series with different Spearman corelation coefficients with the degree vector $\mathbf{d}$. 
			We find that there exist a threshold that the effect reverse. 
			\textbf{f},
			We divide individuals into Hubs (with larger degrees) and Leaves (with smaller degrees).
			We find that if the personal information weights of leaves $\theta_L$ are fixed, increasing the personal information weight of Hubs $\theta_H$ will initially suppress cooperation and then promote it. When the $\theta_L$ is sufficiently large, a greater $\theta_H$ can even perform better than completely ignoring personal information.
			\textbf{g},
			We give a network as an example. 
			The network is composed by two stars and linked by an edge between leaves. 
			So there are two hubs (red circle) and eight leaves (purple circle).
			\textbf{h}, on the network in g, if the average personal information weight $\langle \theta \rangle$ is fixed, increasing $\theta_H$ can initially inhibit cooperation and then promote it.
			When $\langle \theta \rangle$ is large, larger $\theta_H$ can better promote cooperation than the case where $\theta_H=0$.
			\textbf{i}, 
			We consider eight distinct information situation.
			In the first four cases, their $\langle \theta \rangle$ is around 0.2. Among them, a larger $\theta_H$ promotes cooperation the most, while a moderate $\theta_H$ suppresses cooperation. 
			In the last four cases, the cooperation is most inhibited by smallest $\theta_H$ with SCC$=-1$.
		}
		\label{fig: 3}
	\end{figure}

We now turn to investigating the impact of heterogeneous personal information. 
Unlike social information, which only affects the competition rates, personal information influences both the payoff and competition rates (nearly every variable in Eq.~(\ref{eq:competition})). 
This makes analytical analysis extremely challenging.
Nevertheless, we can still examine the effects arising from the coupling between personal information and individual degree.

To investigate the impact of the relationship between personal information sequences and node degree on cooperation, we generate three types of personal information sequences: one that increases linearly, one that decreases linearly, and one that is homogeneous. These sequences are assigned to individuals ordered in descending degree. 
We find that, on heterogeneous networks, a sequence positively correlated with degree promotes cooperation, whereas a negatively correlated sequence suppresses it (Fig.~\ref{fig: 3}a). 
On the contrary, as shown in Fig.~\ref{fig: 3}b, different sequences of personal information have no effect on the evolution of cooperation in regular networks.
These results once again highlight the profound difference between heterogeneous and homogeneous networks.
We also investigate the case where social information and personal information are coupled, as shown in Fig.~\ref{fig: 3}c and d.
We find that the two effects can be combined; for example, the optimal social information sequence together with a personal information sequence that is positively correlated with node degree can jointly promote the emergence of cooperation.

To more comprehensively illustrate the effect of coupling between the personal information sequence and the degree sequence, we generate a range of sequences with different Spearman correlation coefficients (SCC). As shown in Fig.~\ref{fig: 3}e, when the average weight of personal information is low, sequences that are negatively correlated with degree slightly promote cooperation. However, as the average $\theta$ increases, positively correlated sequences begin to promote cooperation more effectively, and this effect becomes more pronounced with larger values of $\theta$.

The above findings imply that if hub individuals rely more heavily on personal information when making decisions, cooperation is more likely to be promoted. 
To more clearly illustrate the role of hub individuals, we fix the personal information weights of low-degree individuals (leaves) and vary the weight assigned to high-degree individuals (hubs). 
As shown in Fig.~\ref{fig: 3}f, when the personal information weights of low-degree leaf individuals are high, increasing the personal information weight of hub nodes $\theta_H$ initially increases the critical threshold $(b/c)^*$ (inhibiting cooperation), but then causes it to decrease (promoting cooperation). Notably, when $\theta_H$ is sufficiently large, its effect in promoting cooperation can even exceed that of the case where $\theta_H = 0$. This contrasts starkly with the earlier conclusion under homogeneous conditions, where smaller $\theta$ always promotes cooperation more strongly \cite{wang2023imitation}.

To more clearly illustrate the impact of hubs, we present an example in Fig.~\ref{fig: 3}g: a network consisting of two stars connected by an edge between two of their leaves.
We plot the curves in Fig.~\ref{fig: 3}h showing how varying the personal information weight of the hubs $\theta_H$ affects cooperation, while keeping the average personal information weight across all nodes $\langle\theta\rangle$ fixed.
As $\theta_H$ increases, cooperation is first suppressed and then promoted. 
The point at which cooperation is most suppressed—corresponding to the maximum value of $(b/c)^*$—may occur either before or after the point where all individuals have equal personal information weights, depending on the value of $\langle\theta\rangle$. 
Moreover, when $\theta_H$ is sufficiently large, its effect in promoting cooperation can even surpass that of the case where $\theta_H = 0$.

We present two groups of examples with different information configurations. In the first group, where $\langle\theta\rangle = 0.2$, cooperation is most suppressed when all individuals have identical personal information weights. 
However, when $\theta_H$ is relatively large, cooperation is promoted. 
Remarkably, further increasing $\theta_H$—even though this raises the average personal information weight $\langle\theta\rangle$—continues to enhance cooperation. 
This is in stark contrast to the conclusions drawn in homogeneous populations, where higher personal information weights typically hinder cooperation.
In the example with $\langle\theta\rangle = 0.4$, we observe a similar pattern: a larger $\theta_H$ promotes cooperation. 
Conversely, reducing $\theta_H$—even though it lowers the average personal information weight—does not lead to further promotion of cooperation.

\subsection*{Information-based preferential attachment}
\begin{figure}[h]
	\centering
	\includegraphics[width=1\textwidth]{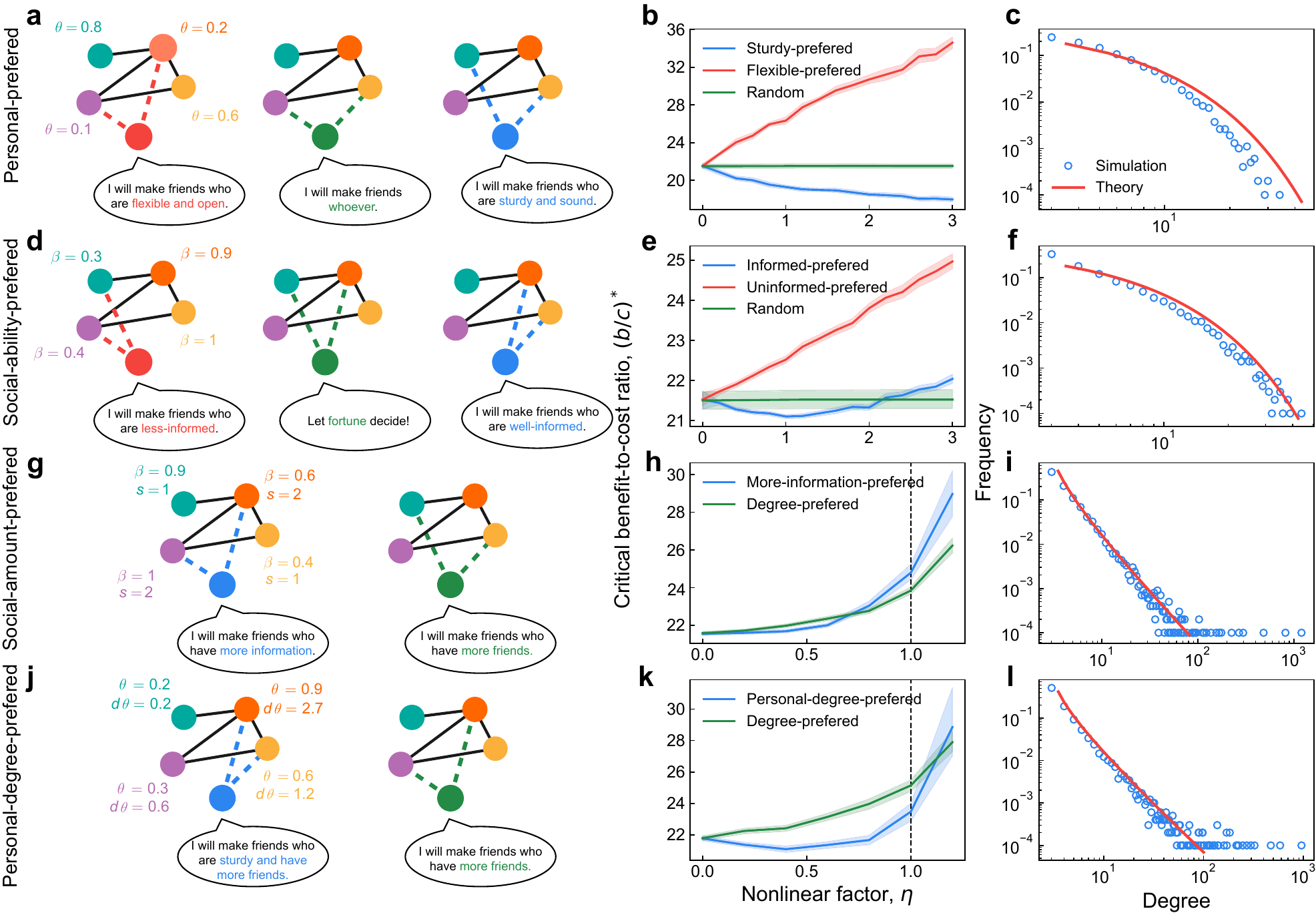}
	
	\caption{
        \textbf{Preferantial attachment based on information.}
		\textbf{a},
		The atractiveness of individual $i$ may be determined solely by the personal information weight $\theta_i$.
		\textbf{b},
		We plot the critical benefit-to-cost ratio, $(b/c)^*$ as a function of nonlinear factor $\eta$.
		Preferring larger $\theta_i$ (Sturdy-preferred) can promote cooperation more than random linking, while preferring smaller $\theta_i$ (Flexible-preferred) inhibit cooperation.
		\textbf{c},
		We plot the degree distribution obtained from simulations alongside the theoretical predictions.
		The distribution decays rapidly as the degree of the individuals increases.
		\textbf{d},
		The atractiveness can also be related to social information $\beta_i$.
		\textbf{e},
		Compared to random linking, as $\eta$ increases, this mechanism initially promotes cooperation and then suppresses it.
		\textbf{f},
		The degree distribution is similar to c.
		\textbf{g},
		Individuals may connot know the social information level $\beta$ of other individuals.
		They may use the obsevation of the amount of social information $s=\lceil \beta_i d_i \rceil$ instead.
		\textbf{h},
		Compared to solely degree-dependent case, cooperation can be promoted when $eta$ is small, and inhibited when it is large.
		\textbf{i},
		The degree distribution in this case is heavy-tailed but not power law distribution.
		\textbf{j},
		The atractiveness may also depends on the combination of $\theta$ and $d$.
		\textbf{k}, \textbf{l}, the effect is similar to h and i.
		For all degree distributions, we take $\eta=1$.
	}
	\label{fig: 4}
\end{figure}

From the results above, we have seen how the distribution of information influences the evolution of cooperation on a given network. 
However, in many populations, once a network has formed, we can no longer dictate the information profile of each individual any more. 
From another perspective, nevertheless, if we regard an individual's information profile as part of their inherent character, this character may itself shape the formation of the network, and ultimately influence how cooperation evolves across the entire population.

An individual's personal information weight $\theta_i$ as a measure of how stubborn or independent they are. 
At the same time, we introduce a new parameter $\beta_i$ to represent the individual's level of social information, which determines their capacity to acquire information from others. 
Hence $s_i = \lceil \beta_i d_i \rceil$, representing an individual's amount of social information as a combined outcome of its degree and its social information level.

All of the above information features may influence network formation. 
For example, a more stubborn individual may attract more connections.
To investigate the such impact, we now introduce an information-based preferential attachment mechanism. 
Suppose that an individual's character (i.e., their information profile) influences how attractive they are to others, denoted by the attractiveness $\alpha_i(\theta_i, \beta_i)$. 
In our model, we start with $m_0 = m + 1$ nodes, which are fully connected. 
Then, we sequentially add new nodes, each with $m$ edges, which are connected to $m$ existing individuals in the network based on their attractiveness. 
The probability that an existing individual $i$ is chosen to receive a link is given by
\begin{equation} \label{eq:PAProb_main}
		\Pi_i=\frac{\alpha_i}{\sum_{j=1}\alpha_j}.
\end{equation}

We consider four typical types of information-based attractiveness. 
The first type assumes that attractiveness depends solely on personal information weight, expressed as $\alpha_i = \phi(\theta_i)^\eta$, where $\eta$ is a nonlinearity factor that captures the marginal effect of variations in $\theta_i$. 
We assume that personal information is uniformly distributed, with $\underline{\theta} \ge 0$ and $\overline{\theta} < 1$ representing the lower and upper bounds of personal information, respectively.
 
As Fig.~\ref{fig: 4}a shows, we examine two specific mappings for $\phi(\theta_i)$ as shown in Fig.~\ref{fig: 4}a: one where $\phi(\theta_i) := \theta_i$, implying that personal influence is positively correlated with attractiveness; and another where $\phi(\theta_i) := \overline{\theta} + \underline{\theta} - \theta_i$, implying a negative correlation between personal influence and attractiveness.
Both cases are compared with the scenario of random connections.
The results shows that when individuals prefer to connect with sturdier partners—those with greater personal information—cooperation is more likely to evolve (Fig.~\ref{fig: 4}b). 
In contrast, when individuals are more attracted to flexible partners—those with lower personal information—cooperation is suppressed. 
Moreover, this effect becomes more pronounced as the difference in attractiveness across varying levels of personal information increases, i.e., with larger values of $\eta$.
We compute the degree distribution of the generated networks, which reads
\begin{equation}
		p(d)=\int_{\underline{\theta}}^{\overline{\theta}} \frac{\int_{\underline{\theta}}^{\overline{\theta}}\phi(x)^\eta dx\,
			e^{1-\frac{ \phi(\theta) ^{-\eta}d }{\left(\overline{\theta}-\underline{\theta}\right) m}\int_{\underline{\theta}}^{\overline{\theta}}\phi(x)^\eta dx}}{m(\overline{\theta}-\underline{\theta})^2}\, d\theta.
	\end{equation}
And the results show a strong agreement between our theoretical predictions and simulation outcomes (Fig.~\ref{fig: 4}c).
Such networks do not exhibit scale-free properties, meaning they do not give rise to a few highly connected hubs.

We also investigate the case where individual $i$'s attractiveness depends on the level of social information $\beta_i$ (Fig.~\ref{fig: 4}d, see SI for details). 
Social information levels are uniformly distributed, with $\underline{\beta} > 0$ and $\overline{\beta} \le 1$ representing the lower and upper bounds of social information, respectively.
If individuals prefer to connect with uninformed partners, cooperation is strongly suppressed (Fig.~\ref{fig: 4}e).
In contrast, if they prefer to connect with informed individuals, cooperation is initially promoted when $\eta$ is small. 
However, as $\eta$ increases, cooperation is ultimately suppressed compared to the case of random connections.
The networks generated under this scenario are similar to those in the previous case (Fig.~\ref{fig: 4}f).

However, individuals may not be able to accurately perceive others' social information levels. 
Instead, they can only observe the amount of social information others possess.
The case where attractiveness is given by $\alpha_i = s_i^{\eta}$ (Fig.~\ref{fig: 4}g) is investigated, where $s_i = \lceil \beta_i d_i \rceil$. 
We compare this with the mechanism based solely on degree preference.
Under such mechanism, a preference for individuals with more social information promotes cooperation only when $\eta$ is relatively small (Fig.~\ref{fig: 4}h). 
In all other cases, this preference suppresses cooperation even more than simply favouring individuals with higher degrees.
The degree distribution of the generated network when $\eta = 1$ is
	\begin{equation}\label{pdSA_mian}
		\begin{aligned}
			p(d)=\frac{\left(\overline{\beta}+\underline{\beta}\right)}{\left(\overline{\beta}-\underline{\beta}\right)}
			\frac{1}{d }
			\left[\mathrm{Ei}\left(\frac{2 \langle\beta\rangle \log \left(\frac{m}{d}\right)}{\underline{\beta}}\right)
			-\mathrm{Ei}\left(\frac{2 \langle\beta\rangle \log \left(\frac{m}{d}\right)}{\overline{\beta}}\right)\right],
		\end{aligned}
	\end{equation}
where $\langle\beta\rangle=(\overline{\beta}+\underline{\beta})/{2}$.
$\mathrm{Ei}(\cdot)$ denotes the exponential integral.
As shown in Fig.~\ref{fig: 4}I, our theoretical results align well with the simulations. 
This is a distinctive heavy-tailed distribution, though it is important to note that it is not a power-law distribution.
When all individuals have the same level of social information, the distribution reduces to a power-law form (see SI for details).

An obvious extension, then, is to consider individual attractiveness as depending on both degree and personal information, given by $\alpha_i = (\theta_i d_i)^{\eta}$ (Fig.~\ref{fig: 4}j).
We find that, compared to a preference for social information amount, a preference for both degree and personal information promotes cooperation over a broader range of conditions (Fig.~\ref{fig: 4}k). 
Moreover, the resulting network exhibits a heavy-tailed degree distribution (Fig.~\ref{fig: 4}l, see SI for details).

\subsection*{Information coevolution}

Beyond individual heterogeneity, another noteworthy point is the heterogeneity introduced by strategies themselves. 
Individuals adopting different strategies may behave differently in their decision-making processes.
This situation can be regarded as a change in the inherent properties of strategies. 
For instance, cooperators may tend to adhere to a benevolent conscience and resist altering their strategies, which leads to $\theta_{\text{C}}>\theta_{\text{D}}$. 
Conversely, defectors may perceive themselves as possessing shrewder strategies and therefore maintain their course ($\theta_{\text{D}}>\theta_{\text{C}}$). 
Meanwhile, the information-gathering capabilities of cooperators and defectors may vary, cooperators might be more popular among their neighbours and thus have easier access to information ($s_{\text{C}}>s_{\text{D}}$).

\begin{figure}[h]
	\centering
	\includegraphics[width=1\textwidth]{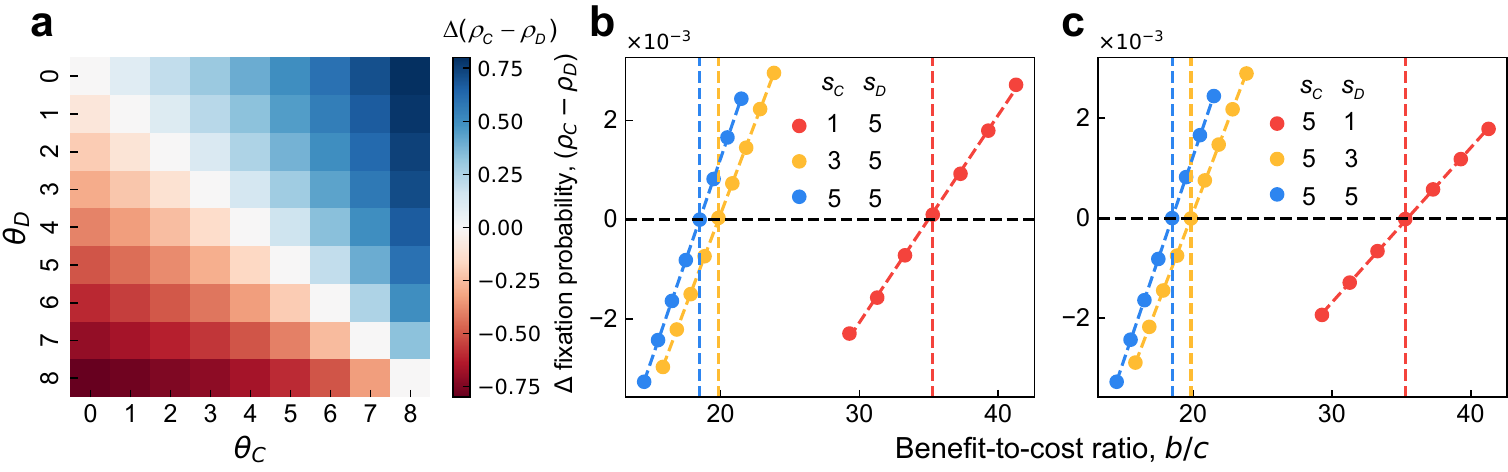}
	
	\caption{
		\textbf{Strategy evolution with coevolution with information.}
        We perform simulations on regular networks.
		\textbf{a},
		We plot the fixation probability difference $\Delta \rho_C-\rho_D$ under different personal information weight for distinct strategies, $\theta_C$ and $\theta_D$.
		The fixation probability of cooperators has a significant advantage when $\theta_C>\theta_D$.
		\textbf{b}, \textbf{c},
		We plot the fixation probability difference $\Delta \rho_C-\rho_D$ as a function of benefit-to-cost ratio $b/c$ under different amount of social information with distinct strategies.
		The role of $s_C$ and $s_D$ are the same.
        For the newtork used, $N=100$, and the degree of all individuals $d=5$.
	}
	\label{fig: 5}
\end{figure}

We separately investigate the coevolution of personal information and social information with strategies.
When cooperators and defectors have different personal information levels, denoted by $\theta_C$ and $\theta_D$ respectively, cooperators are favoured if and only if
\begin{equation}
    \theta_C>\theta_D.
\end{equation}
Moreover, the effect introduced by this gap is highly significant—so much so that it goes beyond the influence typically associated with weak selection. 
In fact, under weak selection, if $\theta_C \neq \theta_D$, then the impact of the game payoffs on the evolution of cooperation becomes almost negligible.
In Fig.~\ref{fig: 5}a, we plot the effect of varying personal information weights of cooperators and defectors on the difference in their fixation probabilities. 
It is clearly observed that when $\theta_C > \theta_D$, the fixation probability of cooperators exceeds that of defectors, and this difference becomes more pronounced as the gap between $\theta_C$ and $\theta_D$ increases.

When cooperators and defectors have the same personal information weight ($\theta_C = \theta_D=\theta$), we can investigate the coevolution of social information and strategy.
Assuming that each individual has a social information amount of $s_{iD}$ when acting as a defector and $s_{iC}$ when acting as a cooperator, the critical threshold at which cooperation is favoured over defection on regular networks is given by
\begin{equation}\label{eq:bcDiffCD_main}
    \left(\frac{b}{c}\right)^*=\frac{4\theta\left(N-1\right) + \left(1-\theta\right)\left[\frac{(s_{\text{C}}-1)d}{s_{\text{C}}(d-1)}+\frac{(s_{\text{D}}-1)d}{s_{\text{D}}(d-1)}\right]\left(N-2\right)}{-4\theta +\left(1-\theta\right)\left[\frac{(s_{\text{C}}-1)d}{s_{\text{C}}(d-1)}+\frac{(s_{\text{D}}-1)d}{s_{\text{D}}(d-1)}\right]\left( N/d-2\right)}.
\end{equation}
From Eq.~(\ref{eq:bcDiffCD_main}), we see that the roles of cooperators' and defectors' social information are symmetric. 
As shown in Fig.~\ref{fig: 5}b and c, we conduct simulations on a regular network where all individuals have a social information amount of $s_C$ when acting as cooperators and $s_D$ when acting as defectors. 
Swapping $s_C$ and $s_D$ does not change the critical threshold $(b/c)^*$ at which cooperation is favoured. 
For both cooperators and defectors, having greater social information facilitates the evolutionary success of cooperation.

\subsection*{Empirical Evidence from Open-Source Collaboration Networks}

Next, we use GitHub event data for empirical validation. GitHub is a natural test bed for the heterogeneous imitation model because it combines a persistent collaboration network with repeated, directed acts of costly contribution and observable social-information sampling. Submitting a pull request, reviewing code, or resolving issues on repositories owned by others is a concrete cooperative act: the contributor pays time and cognitive effort, whereas repository owners and their surrounding communities receive the benefit. Stars, watches, and code-review interactions, in turn, expose developers to visible behavioural cues from other actors. In physical terms, the platform provides a setting in which structure changes slowly, information arrives locally and intermittently, and individual decision cores differ in firmness.

Operationally, we implement this mapping with a two-window design that separates persistent collaboration structure from short-term information acquisition. The structural baseline $T_0$ spans the seven days preceding the analysis day, whereas the behavioural window $T_1$ is the full UTC day \texttt{2025-02-12 00:00--23:59}. The resulting baseline collaboration graph contains 464{,}853 human nodes and 1{,}989{,}883 undirected edges. Hubs are defined as the top 5\% of developers by baseline degree, yielding 23{,}243 analyzable hub ego-networks. For each developer $i$, the baseline neighbour set is $N(i)$ and structural degree is $d_i=|N(i)|$. In $T_1$, the observation set $Obs(i)$ is built from actions that plausibly sample social cues from direct collaborators, namely starring repositories owned by others and submitting pull-request reviews or review comments. Social information is therefore
\[
s_i = |N(i)\cap Obs(i)|,
\]
so the empirical measure remains local and satisfies $s_i\le d_i$ throughout the sample. Personal information weight $\theta_i$ is measured from $T_0$ as the average of cross-owner behavioural consistency and the shrinkage-adjusted rate at which the developer rejects outsider pull requests when acting as a maintainer. Cooperation is measured in $T_1$ as labour contributed on repositories not owned by the focal developer, and local cooperation around a hub is the mean cooperation rate of its $T_0$ neighbours.

\begin{figure}[!h]
\centering
\includegraphics[width=.8\linewidth]{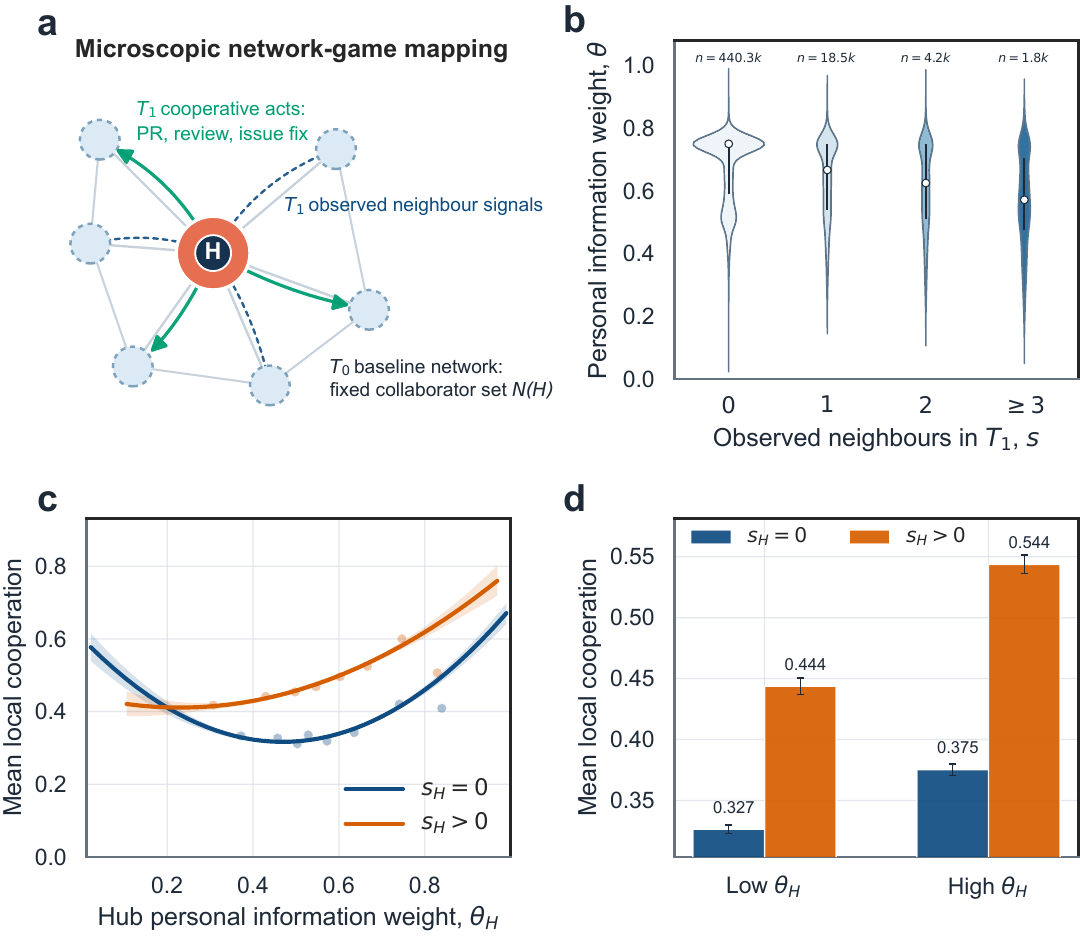}
\caption{\textbf{Empirical evidence from GitHub collaboration networks.} 
\textbf{a}, 
Schematic of the two-window measurement design. The large red node is a hub $H$ identified from the top 5\% of the $T_0$ baseline collaboration network. Its darker inner core denotes larger personal information weight $\theta_H$. Pale dashed circles are the hub's direct collaborators in $T_0$; together they define the structural neighbour set $N(H)$ and hence $d_H=|N(H)|$. Blue dashed arrows indicate neighbour-derived signals actually sampled in the fast window $T_1$, so the empirical social-information variable is $s_H=|N(H)\cap Obs(H)|$. Green solid arrows denote cooperative acts in $T_1$ on repositories owned by others. 
\textbf{b}, 
Distribution of personal information $\theta$ across discrete neighbour-based social-information groups under the main one-day specification. Each violin summarizes developers with $s=0$, $s=1$, $s=2$, or $s\geq 3$; maximum violin width scales with group size, local thickness reflects the within-group density of $\theta$, and labels report the corresponding sample counts. Most developers fall in the zero-information group, but positive-information groups remain large enough to reveal substantial within-group heterogeneity in behavioural firmness. 
\textbf{c}, 
Smoothed relationship between hub personal information and mean cooperation in the hub's neighbourhood. Blue denotes hubs with $s_H=0$ and orange denotes hubs with $s_H>0$. Points are binned means, curves are smooth fits, and shaded bands are 95\% bootstrap confidence intervals. Across most of the support, hubs with positive neighbour-based information sit on a higher cooperation curve, and the curve itself is non-monotonic, first dipping and then rising with $\theta_H$. 
\textbf{d}, 
Mean local cooperation for the four combinations of $s_H=0$ versus $s_H>0$ and low versus high $\theta_H$. Error bars are 95\% confidence intervals. Mean local cooperation rises from 0.327 to 0.375, 0.444, and 0.544, placing the highest cooperation around hubs that combine positive neighbour-based information with high behavioural firmness.}
\label{fig: 6}
\end{figure}

Figure~\ref{fig: 6} links our theoretical framework to empirical collaboration patterns. Figure~\ref{fig: 6}a defines the measurement logic: hubs are identified from a stable collaboration backbone, neighbour-derived information is sampled only from directly connected collaborators in $T_1$, and cooperative acts are quantified as directed labour on external repositories. Figure~\ref{fig: 6}b demonstrates that empirical heterogeneity is inherently two-dimensional. Although most developers sample no direct-neighbour signals within a day, positive-information groups ($s=1$, $s=2$, and $s\geq 3$) remain clearly populated, with personal information weights ($\theta$) broadly distributed across each. This confirms that information heterogeneity is an independent structural feature rather than a simple by-product of activity gradients.

At the hub level, Fig.~\ref{fig: 6}c shows that individuals with positive neighbour-based information lie on a higher cooperation curve across most of the support of $\theta_H$. Crucially, the relationship is non-monotonic: local cooperation first declines and then rises as $\theta_H$ increases, precisely the shape anticipated by our theory (Fig.~\ref{fig: 3}f). 
Figure~\ref{fig: 6}d makes this pattern transparent in grouped form. Mean local cooperation rises from 0.327 for hubs with $s_H=0$ and low $\theta_H$, to 0.375 for hubs with $s_H=0$ and high $\theta_H$, 0.444 for hubs with $s_H>0$ and low $\theta_H$, and finally reaches 0.544 for hubs with $s_H>0$ and high $\theta_H$. 
The most cooperative neighbourhoods are thus concentrated around hubs that combine actual neighbour exposure with a stable decision core, which echoes our results in Fig.~\ref{fig: 2} and \ref{fig: 3}.

Fixed-effects regressions identify which component of this pattern survives the strictest comparative scrutiny. Once hubs are compared within the same owner namespace and differences in baseline degree and short-run activity are absorbed, the robust predictor of a more cooperative neighbourhood is not centrality alone, nor neighbour exposure by itself, but behavioural firmness. Substantively, hubs with higher $\theta_H$ appear to anchor local cooperation more reliably. This aligns with the model's claim that influential actors who rely more heavily on a stable internal decision core are better positioned to sustain cooperative surroundings.

While positive $s_H$ remains a vital descriptive marker for identifying where cooperation is most likely to accumulate (as seen in Fig.~\ref{fig: 6}d), it behaves as an enabling condition rather than a separately isolated driver in the fixed-effects specification. This interpretation is reinforced by a three-day window analysis detailed in the Supplementary Information. Giving developers more time to observe neighbours makes positive $s_H$ more common, yet its coefficient remains statistically imprecise, whereas the effect of $\theta_H$ remains large and stable. The empirical message is therefore asymmetric but sharp: in open-source collaboration, the hubs that most reliably sustain cooperative local environments are not merely well connected or well exposed, but comparatively firm in their own decision core.

\section*{Discussion}

Heterogeneity is among the most pervasive attributes of physical systems, profoundly shaping the complexity and diversity of empirical societies. 
In this study, we have addressed a widely overlooked dimension, dynamical heterogeneity, the notion that each individual inherently possesses distinct characteristics influencing their decision-making processes, here conceptualized through the information used. 
We discover that nodes with high connectivity significantly enhance cooperation when they use more extensive social information and rely more heavily on personal information in their decisions. 
Notably, this promotive effect on cooperation is absent in homogeneous networks, vividly illustrating how heterogeneity can fundamentally transform dynamical outcomes. 
Furthermore, our results reveal the essential role of heterogeneity in the network formation process itself: individuals preferentially connecting with sturdier partners—those relying more substantially on personal information—give rise to networks more conducive to cooperation. 
Additionally, we observe that when cooperators possess greater personal information compared to defectors, cooperation achieves substantial evolutionary advantages; yet intriguingly, social information utilized by cooperators and defectors exerts an equivalent, positive influence on the evolution of cooperation. 
Collectively, these findings underscore the critical role informational heterogeneity plays in shaping not only the dynamics but also the very fabric of the system itself.

Our results clarify the conditions under which specific informational distributions within heterogeneous networks enhance cooperation. 
These insights have profound implications for strategies aimed at regulating collective behaviour to foster cooperative outcomes.
For instance, in social or organisational settings, interventions can strategically provide key individuals (analogous to hubs) with targeted information to promote collective cooperation, such as efficient dissemination of vaccination strategies in public health campaigns \cite{centola2010spread,pastor2002immunization}, or optimised information-sharing policies within organisations to facilitate collaborative innovation and productivity \cite{cross2002people}. 
Our findings offer practical guidelines for designing these informational interventions, highlighting how controlling the distribution of personal and social information can actively steer the collective behaviour of complex networks towards enhanced cooperation.

However, once human or biological systems are established, our capacity to directly modify each individual's informational traits becomes limited. 
An alternative approach involves leveraging the underlying mechanisms of network formation to enhance cooperation. 
We find that preferentially linking to well-informed and sturdy individuals can significantly promote cooperative outcomes. 
This observation aligns with sociological research suggesting that individuals often seek relationships with those who are more informed \cite{granovetter1973strength}, possess higher social status \cite{lin2002social}, or display greater firmness in their beliefs \cite{friedkin1998structural,bikhchandani1992theory}, as such connections can provide access to valuable resources, reliable guidance, and social stability. 
These insights underscore the broader sociological significance of our results, highlighting how the structure and formation of social networks can be instrumental in promoting cooperation.

Additionally, it is noteworthy that cooperation gains a decisive evolutionary advantage whenever cooperators are sturdier—that is, more reliant on personal information—than defectors. Remarkably, this advantage surpasses previously identified mechanisms promoting cooperation under weak selection and is largely independent of specific payoff structures. This finding underscores the critical role of moral mechanisms in promoting cooperation within societies \cite{bowles2011cooperative,haidt2012righteous}. Such morality-based heterogeneity, intrinsically tied to individual strategy, may constitute a fundamental driver behind the emergence and sustainability of cooperation.

The transition from idealized mathematical abstractions to the intricate fabric of real-world interactions represents a critical leap in understanding collective dynamics. 
Our empirical validation within GitHub collaboration networks provides a compelling bridge across this gap, revealing that the interplay between informational access and individual independence is not merely a theoretical construct but a fundamental pillar of large-scale digital ecosystems. 
The observation that hubs with greater ``behavioral firmness'' reliably anchor local cooperation suggests a universal principle of systemic stability: in the face of volatile social cues, the presence of influential actors with a stable decision core prevents the collapse of collective action. 
This discovery transcends the boundaries of traditional game theory, suggesting that the strategic regulation of dynamical heterogeneity could be instrumental in architecting more resilient collaborative systems, from open-source communities to global-scale human-AI collectives. 
By demonstrating that the ``firmness'' of key individuals can counteract the inherent fragility of heterogeneous structures, our work offers a new paradigm for governing the evolution of cooperation in the increasingly complex and information-rich societies of the 21st century.

Although imitation-based update dynamics remain among the most widely employed evolutionary dynamics, numerous alternative dynamics exist in practice, such as aspiration dynamics \cite{zhou2021aspiration} and birth-death updating \cite{ohtsuki2006simple}. 
Exploring heterogeneity within these alternative dynamics constitutes a promising direction for future research. 
In addition, investigating how emerging network structures, such as hypergraphs \cite{alvarez2021evolutionary,wang2026strategy} and temporal networks \cite{li2020evolution}, influence dynamical heterogeneity presents another fascinating avenue for exploration. 
Our study provides a crucial step towards understanding how dynamical heterogeneity shapes the emergence of cooperation in complex systems.

\section*{Acknowledgement}
We thank Aming Li and Lei Zhou for their helpful comments and directions.

\clearpage
\bibliographystyle{naturemag}
\bibliography{ref}

\end{document}